\documentclass[useAMS,usenatbib]{mn2e} 
\usepackage{graphicx,epsfig,psfig} 
\def\spose#1{\hbox to 0pt{#1\hss}} 
\def\simlt{\mathrel{\spose{\lower 3pt\hbox{$\mathchar"218$}} 
\raise 2.0pt\hbox{$\mathchar"13C$}}} 
\def\simgt{\mathrel{\spose{\lower 3pt\hbox{$\mathchar"218$}} 
\raise 2.0pt\hbox{$\mathchar"13E$}}} 
\def\eg{{\rm e.g. }} 
\def\ie{{\rm i.e. }} 
 
\title[The Cold Gas Content of Bulgeless Dwarf Galaxies]{The 
Cold Gas Content of Bulgeless Dwarf Galaxies} 
\author[Pilkington et~al.]{K. Pilkington$^{1,5}$, B.K. Gibson$^{1,5}$, 
F. Calura$^{1}$, A.M. Brooks$^{2}$, L. Mayer$^{3,4}$, 
C.B. Brook$^{1}$,\newauthor
G.S. Stinson$^{1}$, R.J. Thacker$^{5}$, C.G. Few$^{1}$, 
D. Cunnama$^{6}$, and J. Wadsley$^{7}$\\
$^{1}$Jeremiah Horrocks Institute, 
University of Central Lancashire, Preston, PR1~2HE, UK\\ 
$^{2}$Theoretical Astrophysics, Caltech, 1200 E. California Blvd., Pasadena, 
CA, 91125, USA\\
$^{3}$Institut f\"ur Theoretische Physik, University of Z\"urich, Z\"urich, 
Switzerland\\
$^{4}$Department of Physics, Institut f\"ur Astronomie, ETH Z\"urich, Z\"urich, 
Switzerland\\
$^{5}$Department of Astronomy \& Physics, Saint Mary's University, Halifax,
Nova Scotia, B3H~3C3, Canada\\
$^{6}$Physics Department, University of the Western Cape, Cape Town, South  
Africa\\
$^{7}$Department of Physics \& Astronomy, McMaster University, Hamilton, ON, 
L8S~4M1, Canada}

\begin{document} 
\date{Submitted} 
\pagerange{\pageref{firstpage}--\pageref{lastpage}} \pubyear{2010} 
\maketitle 
\label{firstpage} 

\begin{abstract} 
We present an analysis of the neutral hydrogen (HI) properties of a 
fully cosmological hydrodynamical dwarf galaxy, run with varying 
simulation parameters. As reported by \citet{Gov10}, the high 
resolution, high star formation density threshold version of this galaxy 
is the first simulation to result in the successful reproduction of a 
(dwarf) spiral galaxy without any associated stellar bulge. We have set 
out to compare in detail the HI distribution and kinematics of this 
simulated bulgeless disk with what is observed in a sample of nearby 
dwarfs.  To do so, we extracted the radial gas density profiles, 
velocity dispersion (\eg velocity ellipsoid, turbulence), and the power 
spectrum of structure within the cold interstellar medium from the 
simulations.  The highest resolution dwarf, when using a high density 
star formation threshold comparable to densities of giant molecular 
clouds, possesses bulk characteristics consistent with those observed in 
nature, though the cold gas is not as radially extended as that observed 
in nearby dwarfs, resulting in somewhat excessive surface densities. The 
lines-of-sight velocity dispersion radial profiles have values that are 
in good agreement with observed dwarf galaxies, but due to the fact that 
only the streaming velocities of particles are tracked, a correction to 
include the thermal velocities can lead to profiles that are quite flat.  
The ISM power spectra of the simulations appear to possess more power on 
smaller spatial scales than that of the SMC.  We conclude that 
unavoidable limitations remain due to the unresolved physics of star 
formation and feedback within pc-scale molecular clouds.
\end{abstract} 


\section{Introduction} 

A traditional problem plaguing the simulation of
disk galaxies 
\citep[e.g.][and references therein]{Thacker01,som03, Abad03a, 
Gov04, Gov07, Rob04, Bai05,Okam05, San09, Stinson10}, within a 
cosmological context, has been the 
inability to recover successfully the properties of a truly 
``late-type'' disk and, in particular, those with essentially no 
associated stellar bulge, similar to classical galaxies such as M33.

Recent work by \citet{Gov10}, though, has produced what appears to be 
exactly such a bulgeless dwarf, via the imposition of a higher density 
threshold for star formation (100~cm$^{-3}$, as opposed to 
0.1~cm$^{-3}$, as adopted in the aforementioned earlier generations of 
simulations), and mass resolution that allows one to identify individual 
star forming regions.\footnote{The higher star formation density 
threshold can only be applied because the high resolution of the 
simulation, coupled with heating from the UV background, ensures 
fragmentation does not occur at unresolved scales.} The primary dwarf in 
their analysis\footnote{In addition to the supplementary re-simulation of DG1
described in \S~2.} (DG1) forms a shallow central dark matter profile and 
possesses a pure exponential stellar disk of radial scale $r_d$$\sim$1~kpc, 
with a stellar bulge-to-disk ratio $B/D$$\approx$0.04 as determined 
from the $i$ band light profile.

In what follows, we extend this work and 
examine in detail the cold neutral hydrogen 
(HI) gas content of the simulated dwarf DG1 along with its low star 
formation threshold analog, DG1LT, and an updated version of DG1 (called, 
nDG1) which employs high-temperature metal-line cooling 
and enhanced supernova energy feedback to compensate for the additional 
cooling. Our goal is simple: to determine if their HI gas 
properties agree with recent observational data to an equally successful 
degree as the stellar component. Studies such as the The HI Nearby 
Galaxy Survey \citep[THINGS][]{wal08} provide excellent high resolution 
(spectral and spatial) data against which to compare our simulations. 
The gas properties of the simulations are compared directly with several 
of the most recent relevant empirical datasets \citep{Stan99,Tam09, Obr10}, 
in order to assess both their strengths and weaknesses.
  
The cold gas in galaxies is linked directly to underlying star formation 
processes and associated interstellar medium (ISM) physics; any 
successful model of galaxy formation should adopt a holistic approach, 
examining both the gas and star properties in consort.  We describe 
the basic properties of our simulations, before detailing the analyses 
undertaken; we will present results pertaining to the radial 
distribution of cold gas within the disks associated with DG1, DG1LT, 
and nDG1, spatially-resolved velocity dispersion maps of the cold gas, 
and the spatial distribution of power encoded within the structure of 
the ISM. We end with a summary of our findings, discussing both the 
strengths and weaknesses of the current simulations.

\section{method} 

\subsection{Simulations}

We have made use of the recent \citet{Gov10} simulations which produced, 
for the first time, a late-type dwarf spiral with no associated 
stellar bulge. A full description of the simulations' characteristics
is provided by \citet{Gov10}, but for context, it is useful to
summarise their primary traits.

Using the N-body$+$SPH \citep{Mon92} code \textsc{gasoline}, a
low resolution (25~Mpc box, sufficient
to provide realistic torques for these dwarfs), 
dark matter only simulation was used to identify 
3.5$\times$10$^{10}$~M$_\odot$ (virial) halos (with typical
spin paramters $\lambda$=0.05) for potential (high resolution)
re-simulation using a volume renormalisation technique (i.e.,``zoom'' 
simulation).  New initial conditions were then re-constructed 
for the primary target halo (called ``DG1''),
using the relevant low-frequency waves associated with tidal torquing
in the low resolution ``parent'' simulation, but now enhanced with 
higher spatial frequencies generated after tracing the present-day 
particles back to the relevant Lagrangian sub-region within the parent.
The mass distribution was then sampled at higher resolution in the regions
of interest, and more coarsely, further away from the identified halo.
Both DG1 and nDG1 have a force resolution of 86pc, while that of DG1LT
is somewhat lower (116pc); the initial baryonic (dark) particle mass
for DG1 and nDG1 is 3300~M$_\odot$ (16000~M$_\odot$), 
while for DG1LT it is 7800~M$_\odot$ (37000~M$_\odot$).  At $z$=0, 
the $i$-band luminosities of DG1, nDG1, and DG1LT are 
M$_i$=$-$16.5, $-$15.8, and $-$19.1, respectively.

We should re-iterate that each of the three simulations described here
(DG1, DG1LT, and nDG1) use the same dark matter halo / assembly history, 
and
differ primarily only in their treatment of the baryonic physics associated
with star formation - i.e., either supernova energy
feedback efficiency (DG1 vs nDG1) or
star formation density threshold resolution (DG1 vs DG1LT).  DG1
was simulated using a star formation density threshold of 100~cm$^{-3}$, 
typical of the densities encountered in giant molecular clouds, rather 
than the canonical value adopted in earlier simulations 
(0.1~cm$^{-3}$).\footnote{ \textsc{Gasoline} employs an ideal gas 
law equation of state \citep{Wad04}, and the mean molecular 
weight is implicitly solved for and allowed to vary \citep{SWS10}.}
Other than the increased density threshold, two additional 
parameterisations were adopted, within the context of the feedback formalism 
employed: the star formation efficiency ($\epsilon$SF=0.1) and the 
fraction of supernova (SN) energy coupled to the ISM ($\epsilon$SN=0.4). 
The star formation and feedback are modelled as described in 
\citet{Stinson09}.
Without any additional \it ad hoc \rm adjustments, this high density 
threshold led to bulgeless dwarf spirals (akin to the classic prototype, 
M33) with flat (non-centrally concentrated) rotation curves (again, for 
the first time). Alongside our analysis of the high-threshold DG1 
simulation, we provide a parallel analysis of two other simulated 
dwarfs, DG1LT (the lower-threshold analog, which uses the aforementioned 
canonical 0.1~cm$^{-3}$ threshold, and a star formation effiency 
$\epsilon$SF=0.05, with the same initial conditions as that used for 
DG1), and an updated version of DG1, nDG1 (again with the same initial
conditions as DG1 and high 
density threshold of 100~cm$^{-3}$, but now with high-temperature 
metal-line cooling, after \citet{SWS10}, and increased thermal energy 
coupling to the ISM ($\epsilon$SN=1)), in order to better assess the role 
played by star formation threshold and feedback in ``setting'' the gas 
properties of the respective simulations.

To foreshadow the discussion which will follow, perhaps
the most problematic aspect of the current analysis is the uncertain
numerical ``leap-of-faith'' that must be made in associating the typically 
7000$-$8000K SPH gas particles, regardless of their local density
($\sim$0.1$-$100~cm$^{-3}$), with star formation (which in nature occurs
in clouds and cores with temperatures 2$-$3 orders-of-magnitude lower than
this).  Until the effects of cooling by molecular hydrogen are 
incorporated fully within \textsc{Gasoline}, this remains an unavoidable
limitation of our modeling, but fortunately one whose effects are 
known and well-understood.  We return to this point in \S~2.2 and \S~3.3.

\subsection{Analysis}

The cold gas properties of DG1, DG1LT, and nDG1 are compared 
directly with those from comparable dwarfs in The HI Nearby Galaxy 
Survey \citep [THINGS:][] {Tam09}, in addition to the samples of 
\citet{Obr10} and \citet{Stan99}. The bulk properties of DG1 (e.g., 
mass, luminosity, and gas fraction) are consistent with those observed 
in nature \citep[e.g.][]{wal08, VDB01}, and its 
present-day star formation rate ($\sim$0.005~M$_\odot$/yr) and luminosity 
(M$_i$$\approx$$-$16) are (specifically and directly) comparable to 
those of the three dwarfs from \citet{Tam09}, with Holmberg~II (HoII) 
being perhaps the closest direct analog (and, as such, being the 
empirical counterpart to which we will refer DG1 most often). As noted 
earlier, the properties which we derive include the radial extent, the 
velocity dispersion as a function of 
galacto-centric radius, and the power spectrum of the ISM. 

In our work, 
unless otherwise stated, we label ``cold gas'' those SPH particles with 
temperatures less than T$_{\rm max}$=15000~K (after \citet{Stinson06}). 
The bulk of the gas in DG1 (nDG1)
lies near 7000K (9000K), which at face value would appear to be
more appropriate for the warm HI phase of the ISM, rather than
the cold, star-forming, gas, to which we have associated
star formation within the simulation.  However, our cooling, despite
the inclusion of metal-line cooling, is limited
primarily to hydrogen and helium cooling, which can only cool gas
down to these temperatures, and as emphasised in Stinson et al
(2006; \S5.1.1), we are averaging over scales much larger than
individual star forming cores.  The effect of varying this
maximum temperature threshold (T$_{\rm max}$) 
for star formation was examined
in detail by \citet{Stinson06}, to which the reader is referred.  We
can summarise that analysis by stating that provided T$_{\rm max}$ is
chosen to be not too similar in value to that of the mean temperature
of the gas particles, its specific value does not critically affect
star formation (see also, \citet{SWS10}).
Efforts are underway to implement
molecular hydrogen cooling within \textsc{Gasoline}, after which
a quantitative comparison with our results can be undertaken. 

DG1LT, the low density threshold analog to DG1, is analysed in parallel, 
to provide something of a canonical ``control'' sample. As described in 
\citet{Gov10}, the properties of DG1LT (e.g., rotation curve, dark 
matter density profile, bulge-to-disc ratio) are not well-matched to 
those observed in nature, due to the traditional limitations that the 
new suite of simulations were designed to overcome in the first place. 
As a juxtaposition to DG1 though, it is invaluable. The present day star 
formation rate (0.2~M$_\odot$/yr) and luminosity (M$_i$=$-$19.1) are 
much higher than that of DG1 (and the associated stellar mass is 
correspondingly a factor of ten higher), driven (as described by 
\citet{Gov10}) by its adoption of the lower star formation 
threshold (see Fig~1).

\begin{figure}
\centerline{ 
\psfig{file=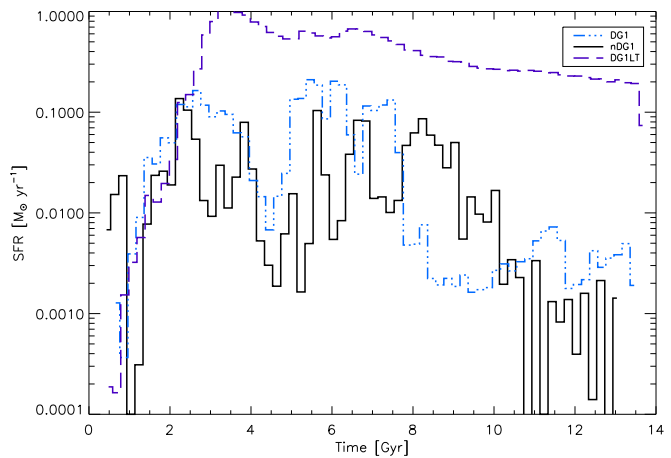,width=9.0cm} 
} 
\caption{The star formation rates of nDG1 (solid line), DG1 
(dot-dashed line), and DG1LT (dashed line). 
Star formation in nDG1 is suppressed overall, 
relative to DG1, but extends $\sim$2~Gyrs beyond the cessation of bulk 
star formation in DG1 (in the range 8$\simlt$$t$$\simlt$10~Gyrs). 
There is intermittent star formation in both 
dwarfs up to the present day. but it has been consistently low for the 
past $\sim$3~Gyrs in nDG1 and $\sim$5~Gyrs in DG1. The star formation 
history of DG1LT is overall considerably higher than its two higher 
density threshold analogs.}
\end{figure} 

For our analysis, we have generated a new variant of DG1 (labelled
nDG1), employing both the same initial conditions and
the higher star formation threshold (100~cm$^{-3}$).  As alluded to
earlier, where nDG1 differs from its predecessors is in its inclusion
of metal-line cooling (following \citet{SWS10}) and a
more efficient coupling of SN thermal energy to the ISM; qualitatively,
we can anticipate this leading to a somewhat more turbulent ISM.
On the whole, the star formation rate of nDG1 is suppressed relative to 
DG1, but extends to lower redshifts (Fig~1, where one can see that
the star formation rate from 8$\simlt$$t$$\simlt$10~Gyr is $\sim$10$\times$
higher in nDG1 than in DG1); its luminosity is, not
surprisingly, somewhat lower than that of DG1 (M$_i$=$-$15.8, as opposed to
M$_i$=$-$16.5), considering
its stellar mass is a factor of two lower 
(M$_\ast$$\approx$2.1$\times$10$^8$~M$_\odot$ vs 
M$_\ast$$\approx$4.4$\times$10$^8$~M$_\odot$).

Zeroth (density), first (velocity), and second (velocity dispersion) 
moment maps of the simulated neutral hydrogen distributions were 
generated using {\textsc{tipsy}}\footnote{\tt 
www-hpcc.astro.washington.edu/tools/tipsy/tipsy.html}, 
after matching the $\sim$40$^\circ$ inclination of the 
dwarfs from the \citet{Tam09} THINGS sample (which, again,
includes HoII, our primary analog against which our simulations will be
compared, as noted in \S~2).
The conversion from ``cold gas'' to ``HI'' within 
{\textsc{ gasoline}} suffices for the purposes 
outlined here; the values derived are close to the values one would predict 
under the assumption of combined photo- and collisional-ionisation equilibrium.
All our results were cross-checked using both cold gas 
and HI moment maps, in addition to further cross-checks undertaken after 
eliminating high column density HI gas for which the conversion from 
cold gas to HI is most insecure.  The results described here are robust 
to these choices, and for expediancy are not discussed further.

Our velocity dispersion analysis made use of the second HI moment map 
derived from the line-of-sight dispersion map produced from viewing the 
DG1, nDG1 and DG1LT simulations with an inclination angle matching that
of HoII.  For the analysis of the distribution of structural ``power'' 
within the cold ISM of the simulations, we again used the zeroth HI moment 
maps and their Fourier Transforms, and compared the inferred power law 
spectra with that derived for the SMC by \citet{Stan99}.

\section{results} 

\subsection{Radial Density Profiles}

We first confirmed independently that the stellar light associated with DG1 
was indeed consistent with a pure exponential of scalelength 
$\sim$1~kpc (\ie bulgeless) disk (akin to the 
Type~I profiles categorised by, for example, \citet{PT06}); 
as shown in the lower panel of
Fig~2, this was the case.  DG1LT also has a radial (stellar) 
scalelength of $\sim$1~kpc, but shows the classical ``problem'' of 
possessing an substantive stellar bulge within the inner kpc
($B/D$$\approx$0.2). 
The stellar disk component of nDG1 is not well-represented by a 
single pure exponential (cf. DG1); instead, its surface density profile
shows a deficit of matter (and light) in the outskirts of the 
stellar disk (beyond a so-called ``break radius'' at 
$\sim$2$-$3~kpc), consistent with the more common
Type~II profiles observed in nature \citep[e.g.][]{PT06,San09}; 
the inner and outer parts of the 
nDG1 stellar disk show radial scalelengths of $\sim$2~kpc and
$\sim$1~kpc, respectively. The bulge-to-disc ratio of nDG1 matches 
formally that of DG1, although it is also readily apparent that the 
surface density (and light) profile of nDG1 shows a high-density 
stellar ``core'', in which $\sim$10$^7$~M$_\odot$ ($\sim$10\% of 
the nDG1 stellar mass, as a whole) is concentrated within the inner 
100~pc.  Importantly, this stellar ``core'' is inconsistent 
with a bulge.  Instead, it consists of a large cluster of stars 
that was formed in the disk during a merger at high-redshift, and 
traveled inward with time so that at $z$=0 it is close to, but not 
located at, the dynamical center of the galaxy (i.e., it can be 
seen to rotate about the galaxy center).

The cold gas of DG1 displays a rapid increase in density within 
$\sim$1kpc.  Exterior to this is an extended disk with an exponential 
scalelength $r_d\sim$6~kpc; the cold gas disk truncates at $\sim$1$r_d$, 
somewhat short of those observed by \citet{Tam09} and \citet{Obr10}, 
where the respective HI disks are traced out to $\sim$2$-$6~$r_d$.  
\citet{Big08} showed that there is an empirical HI upper 
limit encountered in nature - $\Sigma_{HI}$ $\simgt$9~M$_\odot$/pc$^2$.
This upper limit is represented by the horizontal 
line in the upper panel of Figure~2.  
Because we do not yet resolve the microphysics associated with molecular
processes on parsec-scales, one might ascribe some fraction of the cold 
gas in the simulation (particularly that above the upper limit observed 
by \citet{Big08}.) to molecular gas.  Again using the results from 
Bigiel et~al. for the fraction of H$_2$/HI as a function of radius (see 
their Figure 13), we can verify that the high density gas interior to 
1~kpc is consistent with being molecular gas.  In 
particular, $r_{25}$, the isophotal radius corresponding to 
25~mag/arcsec$^2$, is 2.0~kpc for DG1.  Assuming that as much of the 
gas can be ascribed to HI as possible (i.e., the upper limit of 
$\Sigma_{HI}$ = 9~M$_\odot$/pc$^2$), then the results from Bigiel et 
al. suggest that 7.9~M$_\odot$/pc$^2$ would typically be in molecular 
gas at the innermost radius of DG1, dropping to 7.2~M$_\odot$/pc$^2$ 
at 0.8~kpc, and declining radidly to $\simlt$0.1~H$_2$/HI at 2~kpc.  That is, 
while the total amount of gas in DG1 is consistent
with empirical bulk scaling relations, and the gas within 1~kpc is 
consistent with being mostly molecular, the cold gas surface densities 
beyond 1~kpc are too high relative to nature.\footnote{We note that while
this is a sample of one, additional simulations from the same 
suite (e.g., DG2 from \citep{Gov10}) show the same behaviour; we
have chosen to focus only upon DG1, for clarity.} 

It is difficult to interpret the source of this excess gas.  
Perhaps this is gas that should instead be lost from the galaxy in winds?
While it may be tempting to suggest that this gas is overly concentrated, 
comparison of the cold gas scale lengths for these simulated galaxies 
(which has been fit beyond $r_{25}$) to the scalelengths 
beyond $r_{25}$, in the sample of \citet{Big10},
suggests that 
the excess gas in these simulations is actually too extended compared 
to real galaxies.  Alternatively, as discussed below for the case of 
DG1LT, additional star formation in the outskirts of the simulated 
galaxy disks could decrease the surface density of gas (as it goes 
instead into stars).  While \citet{brooks2011} showed that the 
$B$-band scale length of DG1 is comparable to observed dwarf galaxies, 
a factor of 1.5 to 2 increase in size is still allowable to 
be fully consistent with nature.  In fact, preliminary tests of 
molecular cooling and star formation in {\sc Gasoline} suggest that 
the star formation is more extended at $z$=0.  Hence, the addition 
of H$_2$ to these simulations may alleviate the problem of this excess 
gas.

\begin{figure} 
\centerline{ 
\psfig{file=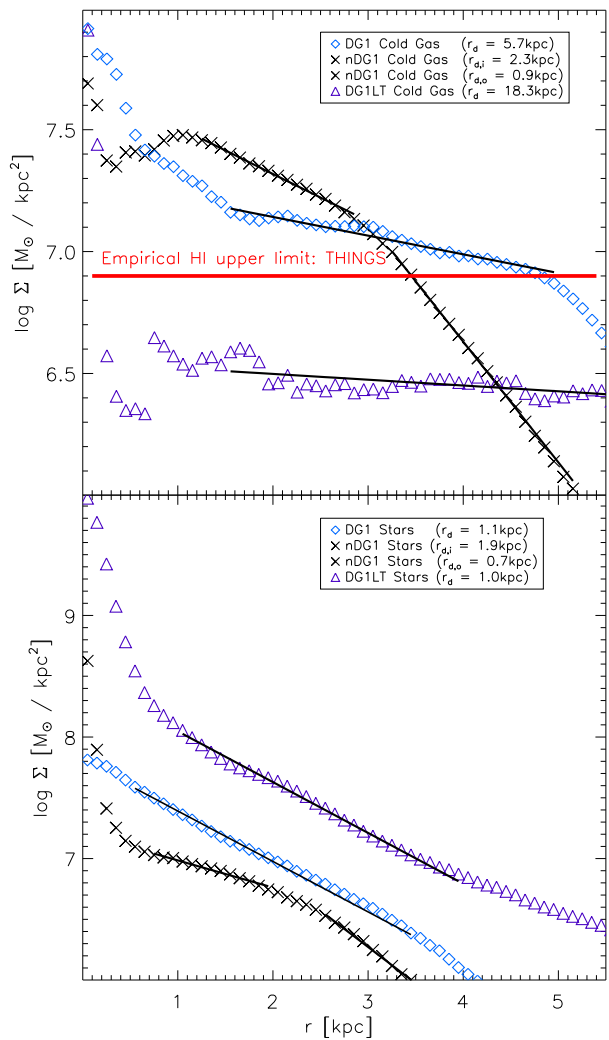, width=9.0cm}
} 
\caption{Radial gas (top) and stellar (bottom) density profiles 
for the simulated dwarfs DG1 (diamonds), DG1LT (triangles), and nDG1 (crosses).
The thick overplotted lines
show the exponential fits to the distributions, from which the noted 
scalelengths were derived. The stellar component of DG1 obeys a 
pure exponential of scalelength $\sim$1~kpc, with no evidence for a 
central bulge, while both nDG1 and DG1LT show central cores. Both the
stellar and cold gas components of nDG1 are best represented by
double exponentials, with a break between the two near $\sim$3~kpc.
The cold gas of DG1 is distributed in a more 
extended exponential disk component of scalelength $\sim$6~kpc, while 
that of DG1LT is $\sim$18~kpc. The horizontal line in the upper
panel corresponds to the empirical upper limit to HI encountered 
in nature, from the THINGS work \citep{Big08}.}
\end{figure} 

As was the case for the stellar light, the disk of nDG1 is better
represented by a ``broken'', or two-component exponential, with 
inner and outer disk scalelengths of $\sim$2~kpc and $\sim$1~kpc,
respectively (with the break occurring near a galactocentric 
radius of $\sim$3~kpc).  The arguments of the previous paragraph
concerning the excess surface density of cold gas in DG1 applies
obviously to nDG1, as well.

Conversely, the cold gas in the disk of DG1LT extends radially to 
$\sim$8~kpc with an essentially flat density profile (formally, with a 
radial scalelength of $\sim$18~kpc -- \ie, the gas disk truncates near 
$\sim$0.5$r_d$ -- again, short of the typical disc in nature, but since
the profile is so flat, the formal exponential ``scalelength'' is 
somewhat ill-defined).  
Like DG1, DG1LT also shows a high density cold gas ``core'' (of mass 
$\sim$2$\times$10$^6$~M$_\odot$), although it is somewhat more extreme, 
in the sense of it being concentrated solely within the inner 
$\sim$100~pc (note that this is 
within twice the force softening length).  Being more
extended, and the gas fraction being an order-of-magnitude
lower (\citet{Gov10}; Tbl~2), it is not surprising that
the cold gas surface density profile of DG1LT is 
consistently a factor of $\sim$3$\times$ lower
than the empirical upper limit derived by \citep{bigiel2008}. 
However, this result should not be interpreted to mean that 
DG1LT is the more realistic version of this galaxy simulation.  
As \citet{Gov10} and \citet{Oh2011} have demonstrated clearly, 
the mass of this galaxy is 
overly concentrated, with a large bulge and peaked inner rotation 
curve that are inconsistent with observed galaxies in the same 
mass range.  

\subsection{Velocity Dispersion}

We next undertook an examination of the velocity dispersion of the HI
disks of DG1, nDG1 and DG1LT, to make a comparison with those observed in
various 
samples of dwarfs in the literature \citep{Cros00, Cros01, Tam09, 
Obr10}. Observations show that independent of present-day star formation 
rate, luminosity, or mass, disks possess a characteristic velocity 
dispersion of $\sim$8$-$10~km/s, rising to $\sim$12$-$15~km/s in the 
inner star-forming regions (\ie within $r_{25}$, the isophotal radius 
corresponding to 25~mag/arcsec$^2$).\footnote{At the 
resolutions at which we are working ($\sim$100~pc), the velocity 
dispersions of the molecular and neutral gas are not 
dramatically different - /citep{Cros00,Cros01}.} A typical 
radial velocity dispersion distribution is shown in Fig~3 for HoII 
(plus signs), from the THINGS sample \citep{Tam09}. 

In addition to the curve for 
HoII, in Figure~3 we also show the corresponding velocity dispersion profiles
(line-of-sight, assuming again a $\sim$40$^\circ$ inclination, similar to that 
of HoII) for DG1 (open diamonds), nDG1 (crosses), and DG1LT (triangles),
derived from the SPH gas particles' \it streaming velocities \rm
(see below, and \citet{VDB02}), and for DG1 (filled diamonds),
taking into account said particles' \it thermal velocities\rm.  
Circular annuli\footnote{Technically, 
elliptical annuli should be used, but our results are not sensitive to 
this choice, at these inclination angles; in addition, we re-measured 
the velocity dispersion profile on the raw THINGS data for HoII using 
circular annuli, to ensure self-consistency with our analysis of the 
simulations.} 
projected on the inclined galaxy were used to set the bins. 

For typical Milky Way-scale simulations, the thermal broadening
component is often neglected, since the `streaming velocity' of
the SPH particle usually dominates over the `thermal component'.
For our simulated dwarfs, this is clearly inadequate, as the 
streaming velocity dispersion can be much smaller than the relevant
thermal velocity dispersion.  To incorporate the latter, 
we follow the procedure outlined by \citet{VDB02}
(\S2.3) and note that the velocity of each particle can be written as
{\bf\it v} = {\bf\it u} $+$ {\bf\it w}, where {\bf\it u} is the 
mean streaming velocity at the location {\bf\it x} and {\bf\it w} is
the particle's random (thermal) velocity.  Because SPH only tracks the
streaming motions of the particles, we make use of the internal energy
of each particle, in order to derive an appropriate random component to
apply to each particle.  In practice, we draw random velocities for each
Cartesian coordinate from a Gaussian of dispersion $\sigma=\sqrt{kT/\mu}$
and add those to each of the coordinates of the streaming motion, 
where $T$ is the temperature of the gas particle
(typically, $\sim$7000$-$9000~K, for our simulations), $k$ is Boltzmann's 
constant, and $\mu$ is the mean molecular weight of the gas.

Without the inclusion of thermal broadening, both DG1 and nDG1 show 
extremely (and unphysically) kinematically cold interstellar media compared 
to DG1LT and, more importantly, dwarfs in nature (compare the crosses and 
open squares of Figure~3 (simulations) with those of the plus symbols 
(observations) for a graphic example of the mismatch between unphysical 
streaming velocity dispersions and those encountered in nature).  
This is not to imply, however, that DG1LT as presented in Figure 3 is 
physical.  First, and most importantly, as already noted
in \S~3.1 and, especially, by \citet{Gov10} and \citet{Oh2011}, the
rotation curve and dynamics of DG1LT are problematic, as is the 
associated significant
overproduction of the stellar bulge.  As can be seen in
Fig~1, DG1LT has a star formation rate two orders of magnitude 
larger than DG1 or nDG1; while this does not impact upon its
consistency with the stellar mass-metallicity, luminosity-metallicity,
or HI gas fraction-luminosity scaling relations, it does worsen
significantly the consistency with the dynamical-to-stellar
mass ratio distribution of \citet{blanton2008}.
This large star formation rate
drives more turbulence, leading to the large streaming velocities for 
this simulation.
We have not included the
thermal component for DG1LT in Fig~3, as doing so would only increase
its velocity dispersion from $\sim$12~km/s to $\sim$14~km/s. 
The inferred line-of-sight velocity dispersion profile for
DG1, after application of the above thermal broadening (which
effectively amounts to a $\sigma$$\sim$7$-$9~km/s broadening of the 
essentially negligible $\sim$1~km/s streaming motions), is represented
by the filled squares in Figure~3.

The characteristic velocity dispersions of the cold gas within DG1 and 
nDG1 are comparable to those encountered in nature ($\sim$8$-$10~km/s - 
\citet{Tam09})
when thermal velocities are considered. 
The thermally broadened 
velocity dispersion profile of DG1 shows a few enhanced features 
(near 0.5r$_{25}$).  These are due to high temperature gas particles 
in and around superbubbles blown by SNe feedback (discussed further 
below and shown in Figure~4).  By design, including a random thermal 
component to the velocity dispersion accentuates these features.  However, 
by chance, the particular timestep we examine here for nDG1 does not 
show any bubbles (though does at previous timesteps), and hence no 
thermal features are introduced into the profile of this 
simulation.  As can be seen from the streaming-only profiles for 
these galaxies, both have slightly higher macroscopic velocity 
dispersions in the inner few hundred parsecs.  However, in DG1 this 
gas is $\sim$35\% hotter than the rest of the disk, while in nDG1 it 
is cooler by a similar factor.  Figure 3 shows that, when this is 
considered in the thermally broadened velocity dispersions, it has 
the effect of maintaining the higher velocity dispersion structure 
in the inner region of DG1, while ``washing out'' the inner structure 
in nDG1.
This result highlights a conundrum in terms of comparing the velocity 
dispersion profiles of these dwarf galaxy simulations to real dwarfs.

A more subtle effect of imposing the random thermal velocity 
perturbation to each particle's streaming motion is that the velocity 
ellipsoid of the cold gas becomes necessarily isotropic, disguising 
any anisotropies that might have been present in the streaming 
motions (i.e., young stars, and the cold gas from which they formed, will 
necessarily have different velocity ellipsoids).  For example, for DG1
(nDG1), the radial, azimuthal, and vertical velocity
dispersions inferred from the cold gas particles' streaming
motions, measured at $\sim$0.5$r_d$, are
$\sigma_r$$\approx$4~km/s ($\sim$6~km/s), 
$\sigma_\phi$$\approx$3~km/s ($\sim$6~km/s), and
$\sigma_z$$\approx$1~km/s ($\sim$2~km/s) -- i.e., 
$\sigma_r$:$\sigma_\phi$:$\sigma_z$$\approx$3:3:1 (anisotropic). 
After thermal broadening,
the derived respective velocity dispersions are
$\sigma_r$$\approx$8.5~km/s ($\sim$10~km/s), 
$\sigma_\phi$$\approx$8~km/s ($\sim$10~km/s), and
$\sigma_z$$\approx$7.5~km/s ($\sim$8.5~km/s) -- i.e., 
$\sigma_r$:$\sigma_\phi$:$\sigma_z$$\approx$1:1:1 (isotropic).
What this means is that 
an unavoidable outcome of our current inability to resolve pc-scale 
molecular heating and cooling processes within the simulations is the 
lack of any significant correlation between velocity dispersion and 
galactocentric radius and/or underlying star formation. 
Until we can resolve densities (and temperatures)
corresponding to the cores of molecular clouds, this apparent 
mismatch between observations and simulations would appear difficult
to avoid.\footnote{It might be tempting to conclude that since
the enhanced feedback did not result in 
a \it significantly \rm higher line-of-sight velocity dispersion, 
this is
consistent with the earlier work of \citet{Dip06} and \citet{PetRup07},
who concluded that supernova 
feedback alone was insufficient to provide 
turbulent heating to the cold ISM in excess of a few km/s; in light of 
the fact that we are not resolving the ISM heating and cooling processes
at pc and sub-pc scales, we feel it premature to draw such a conclusion
from this aspect of our analysis.}

\begin{figure} 
\centerline{ 
\psfig{file=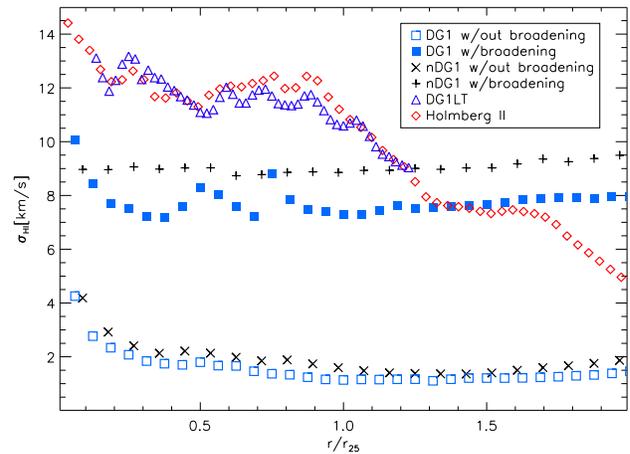,width=9.0cm} 
} 
\caption{Radial behaviour (in units of the B-band $r_{25}$ - \ie the 
isophotal radius corresponding to 25~mag/arcsec$^2$ or, roughly, to the 
extent of the star forming disk) of the HI line-of-sight velocity 
dispersion of the DG1 (open squares), DG1LT (open triangles), and
nDG1 (crosses) simulations, derived from the SPH gas particles'
`streaming velocities' (after \citet{VDB02}), 
in addition to the true HI line-of-sight velocity
dispersion profile for DG1 (filled squares) and nDG1 (plus signs), 
after correcting
the streaming velocities isotropically for their internal thermal
energies. Also shown is
a representative dwarf spiral from the THINGS (Tamburro 
et~al. 2009) sample (HoII: open diamonds).
note: $r_{25}$ is 
2.0~kpc, 5.5~kpc, 1.4~kpc, and 3.3~kpc, respectively, for DG1, DG1LT, 
nDG1, and HoII.}
\end{figure}

\subsection{Power Spectrum and Superbubbles}

\begin{figure*} 
\centerline{ 
\psfig{file=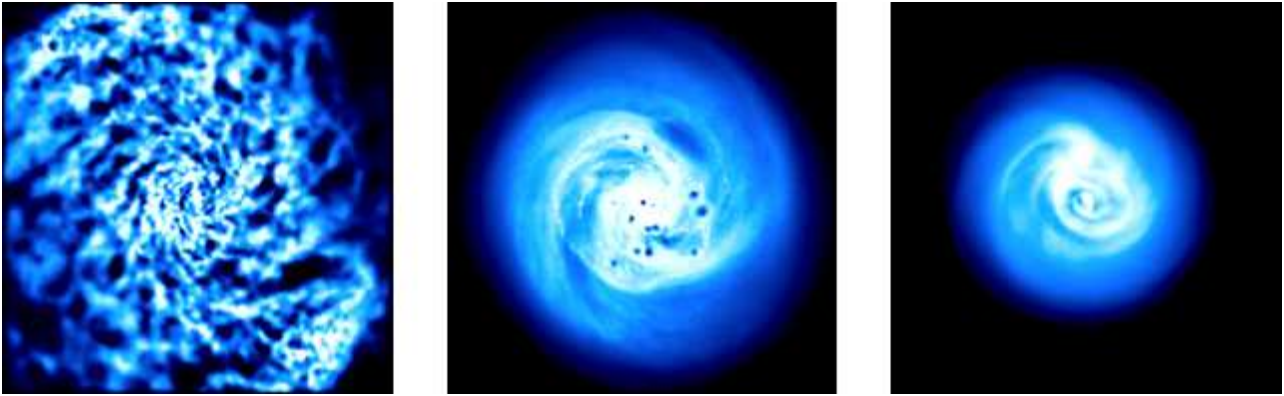,width=17.0cm} 
} 
\caption{Neutral hydrogen (HI) moment zero maps of the three simulations
analysed here - from left to right: DG1LT, DG1, and nDG1.  Each panel
has dimensions 14$\times$14~kpc; a lower column density threshold of 
N(HI)=1$\times$10$^{19}$~cm$^{-2}$ was employed for each map.
}
\label{Pics}
\end{figure*}

\begin{figure} 
\centerline{ 
\psfig{file=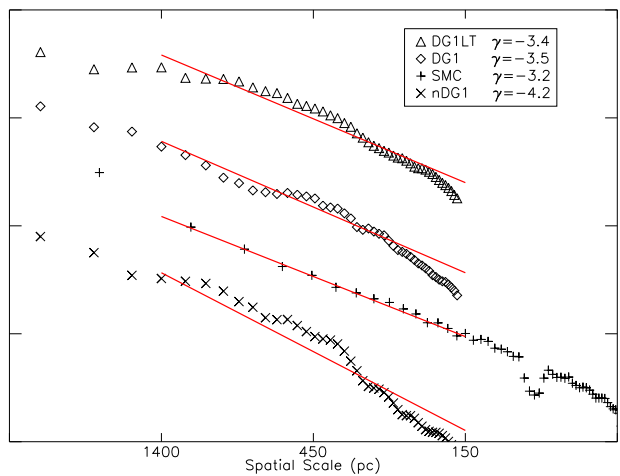,width=8.5cm,bbllx=60bp,bblly=0bp,bburx=504bp,bbury=360bp,clip=}
} 
\caption{Spatial power spectra of the cold ISM of DG1 (diamonds), DG1LT 
(triangles), nDG1 (crosses), and the SMC (plus signs). Power law 
slopes of $-$3.5, 
$-$3.4, $-$4.2 and $-$3.2 are overplotted for DG1, DG1LT, nDG1, and the SMC, 
respectively.  The ``break'' in the SMC power spectrum is due to a 
missing baseline in the \citet{Stan99} ATCA dataset. The
power spectra for the three simulations have been truncated at 
$\sim$2 resolution elements (2$\ast$FWHM of the adopted Gaussian beam:
$\sim$200~pc).}
\label{Power}
\end{figure}

Following \citet{Stan99}, we generated the Fourier Transform of the HI 
moment zero maps of DG1, nDG1, and DG1LT -- each shown in 
Figure~\ref{Pics} at the same spatial scale (14$\times$14~kpc) with the
same limiting HI column density (N(HI)$>$1$\times$10$^{19}$~cm$^{-2}$) -- 
after first convolving the
maps with a 100~pc Gaussian, to mimic the typical beam-smearing
present within THINGS data for HoII \citep{Tam09}.
Circular annuli in Fourier space were then
employed to derive the average power in the structure of the ISM
on different spatial scales.  Figure~\ref{Power} shows the derived power spectra for 
the simulations DG1, nDG1, and DG1LT, and that
for the Small Magellanic Cloud (SMC), re-derived for self-consistency, 
using the HI datacube of \citet{Stan99}. Grossly speaking, the 
distributions can be represented by a power law of the form
$P$$\propto$$k^\gamma$, with $\gamma$=$-$3.5 for DG1,
$\gamma$=$-$3.4 for DG1LT, and $\gamma$=$-$4.2 for nDG1,
and $\gamma$=$-$3.2 
for the SMC (consistent with that found originally by \citet{Stan99}, and consistent with the power spectrum expected when HI density 
fluctuations dominate the ISM structure, rather than turbulent velocity 
fluctuations, which dominate the spectrum when isolating 'thin' velocity 
slices).

There are several points to highlight from Fig~\ref{Power}: 
(i) the SMC shows no evidence for departure from a pure power law, and
hence there does not appear to be any obvious preferred HI cloud size
in nature; (ii) broadly speaking, both DG1 and DG1LT are shallower than
nDG1 (i.e., possess more power on smaller scales, rather than larger, 
relatively speaking); put another way, the enhanced feedback associated
with nDG1 shifts power in the simulated ISM from smaller scales to
larger scales, just as one might expect;  
(iii) each of the simulations shows a greater departure from a pure 
power law, than does the SMC; the most obvious
departure from a power law is perhaps seen in the enhanced power on
$\sim$400$-$500~pc scales in nDG1. This enhanced power corresponds to the
``radial cadence'', or frequency, of the tighly-wound spiral structure 
in the inner few kpcs of the simulation (apparent in the right-most panel
of Figure~4).

Finally, from the present-day moment zero column density map of DG1
(middle panel of Figure~4), we identified 13 SNe-driven
superbubbles in its cold ISM.  
While we do not wish to 
belabour the point when employing such small-number statistics, it is 
re-assuring to note that upon plotting the superbubble size 
distribution, the data were consistent with a power law slope 
between $-$1.5 and $-$2.0 (dependent upon normalisation). Such slopes 
are entirely consistent with those observed in nearby dwarfs 
\citep{OeyCla97}.

\section{Discussion} 

One immediate concern arising from our analysis relates to the 
issue of extracting ``neutral hydrogen'' from the simulations' ``cold gas'' 
(which in some sense consists of both molecular and neutral 
hydrogen).  Because the high-density regions within the simulation
have densities more akin to molecular, rather than neutral,
clouds, it is important to explore the definition of ``neutral''
employed here.\footnote{In large part, this was motivated by the 
fact that in ``column density space'', these high-density
regions possess column densities close to 10$^{22}$~cm$^{-2}$, higher
than those observed in nature; this is a limitation of the 
conversion employed within {\textsc{Gasoline}}.}
To do this, we re-generated HI moment maps, but now restricting the gas 
included to only those particles with densities near the classical value 
of $\sim$0.1~cm$^{-3}$.  As expected, this eliminated the 
unrealistically high neutral hydrogen column densities in the highest 
density regions, but at the expense of leading to vertical density 
profiles that bore little resemblance to the Gaussian profiles observed 
in nature \citep{Obr10}.  Such an extreme ``cut'' to the definition of 
neutral hydrogen also led to a radial profile that bore little 
resemblance to an exponential.  We found no density cut 
which impacted favourably on the observable properties of DG1.  
For these simulations, because
density and temperature are closely correlated in the relevant regime
(T$\simlt$30000~K; $\rho$$\simgt$0.001~cm$^{-3}$), the above
analysis is degenerate to cuts in volume density or temperature.

It is important to note that the primary process responsible for 
driving bulk properties in the simulation is the star formation and 
feedback prescription.  \citet{Gov10} demonstrated that star formation 
had a larger effect on the rotation curve of our simulated galaxy than 
resolution (see their Figure 5).  
The gas properties presented in this paper are 
primarily the result of the star formation prescription, and thus it 
is imperative to use a star formation and feedback prescription that 
is physically motivated.  Until metal-dependent H$_2$ creation and cooling 
is added to the simulations, it is not clear how much HI, as
opposed to H$_2$, should be present in the simulation, 
how it might be distributed as a function of radius, 
and what impact it will have on the resulting disk.


After applying  the physically-motivated $\sim$8~km/s
thermal broadening to the Cartesian coordinates of the SPH particles'
streaming motion, the inferred \it characteristic \rm velocity dispersions for
the cold gas were a reasonable match to those observed in nature (albeit, 
at the unavoidable expense of recovering any correlation between 
velocity dispersion and galactocentric radius and/or global
star formation in the disk, in addition to the imposition of
an isotropic velocity ellipsoid to the cold gas, and the young stars 
which form from this gas).  Beyond the aforementioned issue of the 
lack of a self-consistent treatment of molecular cooling processings 
on sub-parsec scales, one must also
be aware that at the resolutions of these simulations,
we are still missing unresolved star forming regions and
associated turbulence.  The nature of these missing sources is an
active area of debate, but magnetorotational instability (MRI) is one of the 
favoured mechanisms capable of providing a non-negligible amount of
turbulence \citep[e.g.][]{WangAbel09, PionOstr07, MacLow09}

Enhancing the supernovae energy feedback, as was done for simulation
nDG1, at these resolutions, had a marginal impact on the SPH particles'
streaming velocities (at the $\sim$20\% level), which in turn meant 
that its impact on the velocity dispersion profiles was also minimal. 
This is not surprising, as the increased energy deposition was used 
in order to offset the effect of the newly included high-temperature
metal-line cooling.  
Without the inclusion of extra SN energy, the additional cooling that 
comes from metal lines leads to more star formation than in the case 
of DG1.  As shown by \citet{Oh2011}, the stellar mass of DG1 is in 
good agreement with galaxies at similar halo masses, as observed by
THINGS.  If high-temperature metal-line 
cooling had been added with $\epsilon$SN 
held constant, nDG1 would have overproduced stars for galaxies in a 
comparable halo mass range.  
However, the enhanced feedback seems to have steepened the 
spatial power spectrum of the cold ISM of nDG1 relative to DG1, making 
it less consistent with the power spectrum observed for the SMC. 
It is unclear, however, how the power spectrum varies with the 
instantaneous SFR and if this result holds across time. 

Capturing all the relevant ISM physics necessary 
to recover the full spectrum of turbulence sources at pc and sub-pc
scales remains an outstanding challenge.  Despite these limitations, 
the simulated dwarf galaxies presented here have been shown to possess
bulk characteristics consistent with those observed in nature, including 
adherence to scaling relations such as the size-luminosity, size-velocity,
and luminosity-velocity \citep{brooks2011}.
Additionally, the star formation and feedback 
prescription used in these simulations has been shown to result in a
realistic
mass-metallicity relationship as a function of time, and consume gas at a
rate that reproduces the incidence rate and metallicities of both 
QSO-Damped Lyman Alpha (DLA) and GRB-DLA systems \citep{Brooks2007, 
Pontzen2008, Pontzen2009}.  

Hence, it is clear that our simulations remain extremely successful in 
recovering many of the {\it global} optical and dynamical properties of 
realistic bulgeless dwarfs.  That is, although the microphysics of the 
ISM cannot be fully captured at the force resolutions that must be used 
currently in cosmological simulations, this does not largely impact the 
bulk macrophysics such as the rotation curves (stellar and dark matter 
mass profiles), angular momentum content, etc.  On the other hand, we have 
seen that higher resolutions and adoption of more realistic physics for 
star formation leads to simulated galaxies that better reproduce the 
properties of observed galaxies \citep[e.g.,][]{Booth2007, 
RobertsonKravtsov2008, TaskerBryan2008, Saitoh2008, CeverinoKlypin2009, 
Gov10}.  The work presented here highlights paths for future improvement 
in the implementation of ISM physics in cosmological simulations, and provides 
useful tests for reassessment once metal-dependent H$_2$ cooling and star 
formation has been added to {\sc Gasoline} and other cosmological simulation 
codes.

\section*{Acknowledgments} 

We wish to thank the THINGS team for their insights and access to 
their exceptional dataset, in addition to a number of helpful 
discussions with F. Governato and A. Pontzen and comments 
from the referee which led to a vastly improved 
manuscript. BKG and CBB acknowledge the support of the 
UK's Science \& Technology Facilities Council (ST/F002432/1). 
KP acknowledges the support of STFC through its
PhD Studentship programme (ST/F007701/1). 
RJT acknowledges support from NSERC, CFI, the
CRC program, and NSRIT. KP and BKG acknowledge visitor support
from Saint Mary's University.
We thank the DEISA consortium, co-funded through
EU FP6 project RI-031513 and the FP7 project RI-222919,
for support within the DEISA Extreme Computing Initative, and the 
UK's National Cosmology Supercomputer (COSMOS), NASA's Advanced 
Supercomputing Division, TeraGrid, the Arctic Region Supercomputing 
Center, and the University of Central Lancashire's High
Performance Computing Facility.

\bibliographystyle{mn2e} 
\bibliography{DG1_Paper} 

\label{lastpage}

\end{document}